\documentclass[seceq]{ptptex}
\notypesetlogo
\usepackage{wrapft,graphicx}
\title{New Constraint of Clustering for AMD and Its Application\\to the Study of $2\alpha$-\ncl{12}{C} Structure of \ncl{20}{Ne}}
\author{Yasutaka \textsc{Taniguchi}, Masaaki \textsc{Kimura}$^*$ and Hisashi \textsc{Horiuchi}}
\inst{Department of Physics, Kyoto University, Kyoto 606-8502, Japan\\
$^*$The Institute of Physical and Chemical Research(RIKEN), Wako 351-0198, Japan}
\abst{A new constraint of clustering for the AMD calculation is proposed. 
This new constraint gives us large improvement in studying the cluster 
structure by AMD which sometimes meets difficulty in giving rise to some 
specific cluster configurations.  The usefulness of this new constraint 
is verified by applying it to the the study of the third $K^\pi = 0^+$ band 
of $^{20}$Ne which has been discussed to have $2\alpha$-\ncl{12}{C} 
structure.  This band has not been easy even to construct by AMD. 
We see that the AMD+GCM calculation by the use of the new constraint 
gives rise to the third $K^\pi = 0^+$ band which contains the 
$2\alpha$-\ncl{12}{C} structure as an important component.}
\begin{document}
\newcommand{\ncl}[2]{\mbox{$ ^{#1}{\rm #2}$}}
\newcommand{\bra}[1]{\mbox{$\left< #1 \right| $}}
\newcommand{\ket}[1]{\mbox{$\left| #1 \right> $}}
\newcommand{\vect}[1]{\mbox{\boldmath $#1$}}
\newcommand{\norm}[1]{\mbox{$\bra{ #1 }\left.\! #1 \right>$}}
\newcommand{\expec}[1]{\mbox{$\left< #1 \right>$}}
\newcommand{\inpro}[2]{\mbox{$\bra{ #1 }\left.\! #2 \right>$}}
\maketitle
\section{Introduction}
\label{sec:intro}
For the study of the coexistence of the cluster structure and the 
 mean-field-type structure in nuclei, the AMD (antisymmetrized molecular 
 dynamics) method has proved to present us with a powerful approach 
 for both stable and unstable nuclei.\cite{KKH,kimura}  
 AMD is a kind of {\it ab initio} theory which can describe any 
 kinds of nuclear structure but does not require any 
 model assumptions on the nuclear structure such 
 as the assumption of the formation of the mean-field-type structure and the 
 assumption of the existence of any clusters.  One of the frequently used 
 processes of AMD study of nuclear structure is the constrained AMD 
 calculation in which we calculate the parity-projected energy surface as a 
 function of some constraint parameter like in the constrained Hartree-Fock 
 calculation.  As the constraint parameter we usually use the quadrupole 
 deformation parameter $\beta$.  If the system prefers to have a cluster 
 state, it usually appears as the energy minimum point on the 
 parity-projected energy surface as a function of the $\beta$ parameter, 
 which we abbreviate as the $\beta$-constraint surface hereafter.  
 However, the use of the $\beta$-constraint surface can not be always 
 a good way for the study of the cluster structure.  
 It is because the quadrupole deformation parameter is not a direct quantity 
 to characterize the clustering structure but is a good quantity to 
 characterize the mean-field-type structure.  The minimum energy state for a 
 given value of $\beta$ tends to have more mean-field-type character than 
 cluster structure character.  Therefore the states on the $\beta$-constraint 
 surface tend to underestimate the clustering character.  The state 
 having more prominent clustering character is often located higher above 
 the state on the $\beta$-constraint surface at each value of $\beta$.  

 The inappropriateness of the constraint parameter $\beta$ is sometimes 
 rather serious because some kind of cluster structure which is reported to 
 exist in experiments can not be easily obtained on the $\beta$-constraint 
 surface.  One such example is the $K^\pi = 0^-$ band which has 
 $\alpha$-$^{40}$Ca structure\cite{yamaya,ohkubo}.  As is reported in 
 Ref.\citen{Ti44}, this $K^\pi = 0^-$ band with $\alpha$-$^{40}$Ca 
 structure is not easy to obtain on the $\beta$-constraint surface 
 of AMD and hence in Ref.\citen{Ti44} it was constructed by explicitly 
 using the $\alpha$-$^{40}$Ca Brink wave function as the 
 initial state for the frictional cooling process of AMD which is an energy 
 variation process in AMD analogous to the imaginary time method in the 
 Hartree-Fock approach.  Another example is the third $K^\pi = 0^+$ band in 
 $^{20}$Ne which is strongly suggested by experiments to have 
 $2\alpha$-\ncl{12}{C} structure since the band member states are strongly 
 populated by the $^8$Be transfer reaction on $^{12}$C.\cite{transfer}  
 The assignment of the $2\alpha$-\ncl{12}{C} structure to this third 
 $K^\pi = 0^+$ band was supported by the semi-microscopic 
 calculation of the ($^{12}$C-$^8$Be) + 
 ($\alpha$-$^{16}$O) coupled channel OCM (orthogonality condition 
 model).\cite{fujiwara,supple} However, as is seen in 
 Ref.\citen{kimura}, this third $K^\pi = 0^+$ band of $^{20}$Ne could 
 not be shown even to exist on the $\beta$-constraint surface of AMD.
Hartree-Fock calculations have also never 
 shown clearly the existence of the third $K^\pi = 0^+$ band of $^{20}$Ne. 
  
 It is clear that the above-mentioned deficiency of the $\beta$-constraint 
 surface for the proper treatment of the clustering problem is due to the 
 inappropriateness of the constraint parameter $\beta$ for the study of the 
 clustering.  We should invent a new constraint parameter which is more 
 suitable for treating the clustering character and should calculate 
 the parity-projected energy surface as a function of the new constraint 
 parameter.  The purpose of this paper is to propose a new constraint for 
 AMD which is suitable for treating the clustering problem and to verify its 
 usefulness by applying it to the study of the third $K^\pi = 0^+$ band of 
 $^{20}$Ne.  We will see that we succeed in reproducing the existence 
 of the third $K^\pi = 0^+$ band of $^{20}$Ne.  We will also see that 
 this calculated band contains the $2\alpha$-\ncl{12}{C} structure as an 
 important component. 

 In these days, the study of the coexistence of the cluster structure 
 and the mean-field-type structure has become a rather hot issue for 
 nuclei around the mass number 40.  In this region of nuclei, recent 
 experimental studies have assigned many superdeformed rotational bands 
 which have the structure of many-particle many-hole 
 configuration.\cite{AR36,CA40,TI44exp}  Many of the low spin states 
 of these superdeformed bands have long been known to exist\cite{simps} 
 and their natures have been understood often in relation with the 
 clustering\cite{sakuda} or quarteting\cite{quart} correlations.  
 Therefore we are now confronting the situation which requires us to 
 clarify the relation between the mean-field-type structure with 
 superdeformation and the cluster structure with large deformation. 
 We expect that our new constraint will play a powerful role in the 
 study of the coexistence of the cluster structure and the 
 mean-field-type structure in nuclei around the mass number 40. 

 The content of this paper is as follows. In the next section, \S\ref{sec:frame}, 
 we explain briefly the formulation of the deformed-basis 
 AMD\cite{kimura,KKH,KH} and in \S\ref{sec:d-constraint} we introduce and explain 
 a new constraint for clustering which we call the $d$-constraint. 
 In \S\ref{sec:wave_functions}, we prepare the basis wave functions to be used in our 
 present AMD+GCM (generator coordinate method) calculation for $^{20}$Ne.  
 The basis wave functions consist of those constructed with the 
 $d$-constraint in addition to those constructed with the usual 
 $\beta$-constraint.  Two types of cluster configurations are 
 adopted in constructing the basis wave functions with the 
 $d$-constraint; one is the $\alpha$-$^{16}$O type and the other 
 is the $2\alpha$-\ncl{12}{C} type.  In \S\ref{sec:results}, we 
 discuss the results of the GCM calculation.  We will see that our GCM 
 calculation gives rise to the third $K^\pi = 0^+$ band which contains 
 the $2\alpha$-\ncl{12}{C} structure as an important component. 
 Finally in \S\ref{sec:summary}, we give summary.
\section{Framework of Deformed-basis AMD}
\label{sec:frame}
\subsection{Wave Function and Hamiltonian}
In deformed-basis AMD, the intrinsic wave function of the system with mass $A$ is given by a Slater determinant of single-particle wave packets,
\begin{eqnarray}
{\rm\Phi_{int}} &=& \frac{1}{\sqrt{A!}}{\cal A}\{ \varphi_1,\varphi_2,\cdots,\varphi_A\},\label{eq:AMDwf}\\
\varphi_i &=& \phi_i(\vect{r})\chi_i\xi_i,
\end{eqnarray}
where $\varphi_i$ is the $i$th single-particle wave packet consisting of spatial $\phi_i$, spin $\chi_i$ and isospin $\xi_i$ parts. Deformed-basis AMD employs the triaxially deformed Gaussian centered at $\vect{Z}_i$ as the spatial part of the single-particle wave packet:
\begin{eqnarray}
\phi_i(\vect{r}) &\propto& \exp\left[ -\sum_{\sigma = x,y,z} \nu_\sigma\left(r_\sigma - \frac{Z_{i\sigma}}{\sqrt{\nu_\sigma}}\right)^2\right],\nonumber \\
\chi_i &=& \alpha_i\chi_\uparrow + \beta_i\chi_\downarrow,\  |\alpha_i|^2+|\beta_i|^2 = 1,\nonumber \\
\xi_i &=& {\rm proton\  or\  nertron}.
\end{eqnarray}
Here, the complex number parameter $\vect{Z}_i$ which represents the centroids of the Gaussian in phase space takes an independent value for each nucleon. The width parameters $\nu_x,\nu_y$ and $\nu_z$ are real number parameters and take independent values for each direction, but are common to all nucleons. Spin part $\chi_i$ is parametrized by $\alpha_i$ and $\beta_i$ and isospin part $\xi_i$ is fixed to up(proton) or down(neutron). $\vect{Z}_i,\nu_x,\nu_y,\nu_z$ and $\alpha_i,\beta_i$ are the variational parameters and are optimized by the method of frictional cooling explained in the next subsubsection. As the variational wave function, we employ the parity projected wave function:
\begin{equation}
{\rm \Phi}^\pm = P^\pm {\rm \Phi}_{\rm int} = \frac{1\pm P_x}{2}{\rm \Phi}_{\rm int},\label{eq:AMDwf_parity}
\end{equation}
here $P_x$ is the parity operator and ${\rm \Phi}_{\rm int}$ is the intrinsic wave function given in Eq.(\ref{eq:AMDwf}).

The Hamiltonian used in this study is as follows;
\begin{equation}
\hat{H} = \hat{T} + \hat{V}_{\rm N} + \hat{V}_{\rm C} - \hat{T}_{\rm G},
\end{equation}
where $\hat{T}$ and $\hat{T}_{\rm G}$ are the kinetic energy and the energy of the center of mass motion, respectively. We have used the Gogny D1S force as an effective nuclear force $\hat{V}_{\rm N}$. Coulomb force $\hat{V}_{\rm C}$ is approximated by the sum of seven Gaussians.

\subsection{Energy Variation, Angular Momentum Projection and Generator Coordinate Method}
\label{sec:energy_variation}
We perform the variational calculation and optimize the variational parameters included in the trial wave function Eq.(\ref{eq:AMDwf_parity}) to find the state that minimizes the energy of the system $E^\pm$;
\begin{equation}
E^\pm = \frac{\bra{{\rm\Phi^\pm}}\hat{\cal{H}}\ket{\rm\Phi^\pm}}{\norm{\rm\Phi^\pm}},\  \hat{\cal{H}} = \hat{H} + \hat{V}_{\rm cnst}. \label{eq:E_pm}
\end{equation}
We add the constraint potential $\hat{V}_{\rm cnst}$ to the Hamiltonian $\hat{H}$ to obtain the minimum energy state under the optional constraint condition. In this study, we constrain matter quadrupole deformation and impose a new constraint on the distance between quasi-clusters, by employing the potential $\hat{V}_{\rm cnst} = v^\beta_{\rm cnst}(\expec{\beta^2}-\beta_0^2)^2\   {\rm and}\   v^d_{\rm cnst}(\expec{d^2}-d_0^2)^2$ and we obtain the optimized wave function ${\rm\Phi}^\pm_\beta(\beta_0) = P^\pm {\rm\Phi}_{\rm int}^\beta(\beta_0)\  {\rm and}\  {\rm\Phi}^\pm_d(d_0) = P^\pm {\rm\Phi}_{\rm int}^d(d_0)$. As for the distance between quasi-clusters, we will explain it in the next subsection. The evaluation of the quadrupole deformation parameter $\beta$ is explained in Ref.\citen{Dote}. At the end of the variational calculation, the expectation value of $\hat{V}_{\rm cnst}$ should be zero in principle, and in the case of $\beta$-constraint we confirm that it is less than 100 eV. However, in the case of the constraint on the distance between quasi-clusters, the resultant $\expec{d^2}$ value sometimes differs slightly from the aimed value $\expec{d_0^2}$, and hence the expectation value of $\hat{V}_{\rm cnst}$, $\expec{\hat{V}_{\rm cnst}}$, becomes non-negligible. In such a case, we of course subtract $\expec{\hat{V}_{\rm cnst}}$ from $E^\pm$ of Eq.(\ref{eq:E_pm}) in calculating the energy of the system.

Energy variation with the AMD wave function is performed by the frictional cooling method. The reader is referred to Refs.\citen{Ono,Enyo} for a more detailed explanation. The time development equation for the complex number parameters $\vect{Z}_i,\alpha_i$ and $\beta_i$ is as follows;
\begin{equation}
\frac{dX_i}{dt} = \frac{\mu}{\hbar}\frac{\partial E^\pm}{\partial X_i^*},\  (i = 1,2,\cdots,A)
\end{equation}
and for the real number parameters $\nu_x,\nu_y$ and $\nu_z$;
\begin{equation}
\frac{d\nu_\sigma}{dt} = \frac{\mu'}{\hbar}\frac{\partial E^\pm}{\partial\nu_\sigma},\  (\sigma = x,y,z)
\end{equation}
Here $X_i$ is $\vect{Z}_i,\alpha_i$ or $\beta_i$. $\mu$ and $\mu'$ are arbitrary negative real numbers. It is easy to show that the energy of the system decreases as time develops, and after sufficient time steps we obtain the minimum energy state.

From the optimized wave function, we project out the eigenstate of the total angular momentum $J$,
\begin{eqnarray}
{{\rm\Phi}_\beta}^{J\pm}_{MK}(\beta 0) &=& P^J_{MK}{\rm\Phi}^\pm_\beta(\beta_0),\\
{{\rm\Phi}_d}^{J\pm}_{MK}(d_0) &=& P^J_{MK}{\rm\Phi}^\pm_d(d_0).
\end{eqnarray}
Here $P^J_{MK}$ is the total angular momentum projector. The integrals over the three Euler angles included in the $P^J_{MK}$ are evaluated by numerical integration.

Furthermore, we superpose the wave functions ${{\rm\Phi}_\beta}^J_{MK}$ and ${{\rm\Phi}_d}^J_{MK}$ which have the same parity and angular momentum but have different values of deformation parameters $\beta_0$ and distance parameters $d_0$ and $K$. Thus the final wave function of the system becomes as follows:
\begin{eqnarray}
{\rm\Phi}^{J\pm}_n &=& c_n{{\rm\Phi}_\beta}^{J\pm}_{MK}(\beta_0) + c'_n{{\rm\Phi}_\beta}^{J\pm}_{MK'}(\beta'_0) + \cdots \nonumber \\
&&\  + d_n{{\rm\Phi}_d}^{J\pm}_{MK''}(d_0) + d'_n{{\rm\Phi}_d}^{J\pm}_{MK'''}(d'_0) + \cdots,
\end{eqnarray}
where quantum numbers other than total angular momentum and parity are represented by $n$. The coefficients $c_n,c'_n,\cdots,d_n,d'_n,\cdots$ are determined by the Hill-Wheeler equation,
\begin{equation}
\delta\left(\bra{{\rm\Phi}^{J\pm}_n}\hat{H}\ket{{\rm\Phi}^{J\pm}_n} - \epsilon_n\norm{{\rm\Phi}^{J\pm}_n}\right) = 0.
\end{equation}

\section{New Constraint ($d$-constraint)}
\label{sec:d-constraint}
In this paper, we propose a new constraint for creating the wave functions of cluster structure in  AMD. This constraint is imposed on the distance between the centers of mass of quasi-clusters. The meaning of ``quasi-cluster'' will be soon explained below.  We call this constraint {\it $d$-constraint}.

At first, we decide proton number and neutron number of each quasi-cluster. Next, we make a numbering of nucleons to fix which nucleon belongs to which quasi-cluster. By the ``quasi-cluster'', we simply mean a set of nucleons which consists of given number of neutrons and protons. We do not impose any other properties to the quasi-cluster such as the spatial distributions of nucleons of the quasi-cluster. In this paper, we have studied the $\alpha$-\ncl{16}{O} and $2\alpha$-\ncl{12}{C} structures of \ncl{20}{Ne}. Therefore we treat quasi-$\alpha$ cluster and quasi-\ncl{12}{C} cluster and quasi-\ncl{16}{O} cluster in \ncl{20}{Ne}.

The center of mass of quasi-cluster $\rm C_n$, $\vect{R}_{\rm n}$, is defined in the following way:
\begin{equation}
R_{{\rm n}\sigma} = \frac{1}{A_{\rm n}}\sum_{i\in {\rm C_n}}\frac{{\rm Re} Z_{i\sigma}}{\sqrt{\nu_\sigma}}\ \ \   (\sigma = x,y,z), \label{eq:c_of_qc}
\end{equation}
where $A_{\rm n}$ is the mass number of quasi-cluster $\rm C_n$, and $i\in {\rm C_n}$ means $i$th nucleon is contained in the quasi-cluster $\rm C_n$. It is to be noted that the $\sigma (= x,y,z)$ component of the spatial center of one particle wave function, $\varphi_i$, is $\frac{{\rm Re} Z_{i\sigma}}{\sqrt{\nu_\sigma}}$. We define the distance $d$ between quasi-clusters $\rm C_n$ and $\rm C_m$, as
\begin{equation}
\expec{d^2} = \left| \vect{R}_{\rm n} - \vect{R}_{\rm m} \right| ^2.
\end{equation}

If the system favors to have clusters $\rm C_n$ and $\rm C_m$ with the mutual distance $d=d_0$, we will obtain, after the frictional cooling, the wave function in which $A_{\rm n}$ nucleons gather closely around $\vect{R}_{\rm n}$ and $A_{\rm m}$ nucleons gather closely around $\vect{R}_{\rm m}$ where $\left| \vect{R}_{\rm n} - \vect{R}_{\rm m}\right| = d_0$, respectively. On the contrary, if the system does not favor to have clusters $\rm C_n$ and $\rm C_m$ with $d=d_0$, it can happen that $A_{\rm n}$ nucleons and $A_{\rm m}$ nucleons do not gather closely around $\vect{R}_{\rm n}$ and $\vect{R}_{\rm m}$, respectively, where $\left| \vect{R}_{\rm n} - \vect{R}_{\rm m}\right| = d_0$. Furthermore, even when $A_{\rm n}$ and $A_{\rm m}$ nucleons gather closely around $\vect{R}_{\rm n}$ and $\vect{R}_{\rm m}$ with $\left| \vect{R}_{\rm n} - \vect{R}_{\rm m}\right| = d_0$, respectively, these $A_{\rm n}$ and $A_{\rm m}$ nucleons do not necessarily form the ground state configurations of the clusters $\rm C_n$ and $\rm C_m$, but form in general the distorted configurations from the ground state configurations. In summary, the use of $d$-constraint does not necessarily mean to result in the construction of the $\rm C_n$-$\rm C_m$ cluster structure, and even when the $\rm C_n$-$\rm C_m$ structure is formed, the wave function created by $d$-constraint is not necessarily the same as the Brink model wave function as usual cluster models. In the usual Brink model, nucleon wave packets of a cluster is at the same place, which means that the usual Brink model can only describe the cluster at approximate ground state.

\begin{wrapfigure}{l}{8cm}
\centerline{\includegraphics[width=7.5cm]{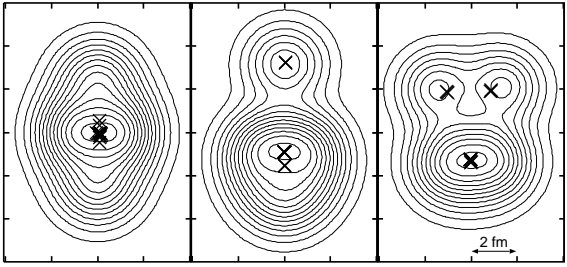}}
\caption{The density distributions of mean-field-type structure($\beta$-constraint), ``$\alpha$-\ncl{16}{O}'' structure($d$-constraint) and ``$2\alpha$-\ncl{12}{C}'' structure($d$-constraint), from left to right: $\times$ are the centroids of the wave packets.}
\label{fig:density_distributions}
\end{wrapfigure}
As examples of calculating 2-body and 3-body cluster wave functions, we discuss ``$\alpha$-\ncl{16}{O}'' and ``$2\alpha$-\ncl{12}{C}'' wave functions which we use later in this paper(Fig.\ref{fig:density_distributions}). The central and right figures are those calculated by $d$-constraint. In getting the ``$\alpha$-\ncl{16}{O}'' wave function of the central figure, the constraint distance $d$ between quasi-$\alpha$ and \ncl{16}{O} clusters is set to $d=4.5{\rm\  fm}$, while in getting the ``$2\alpha$-\ncl{12}{C}'' wave function of the right figure, we constrained the distance $d_i\  (i=1,2)$ between the $i$th quasi-$\alpha$ cluster and the quasi-\ncl{12}{C} cluster to be $d_1=d_2=3.4{\ \rm fm}$, and did not impose any constraint on the distance between two quasi-$\alpha$ clusters. We see that the nucleons are obviously clustering.

In the case of ``$\alpha$-\ncl{16}{O}'' wave functions, when the constrained distance is long, two quasi-clusters go nearly to their ground states. This is comparable to the Brink model. When the constrained distance is short, nucleon wave packets are deformed and wave function is similar to a mean-field-type state. This is different from the Brink model.

\begin{wrapfigure}{r}{7.5cm}
\centerline{\includegraphics[width=7cm]{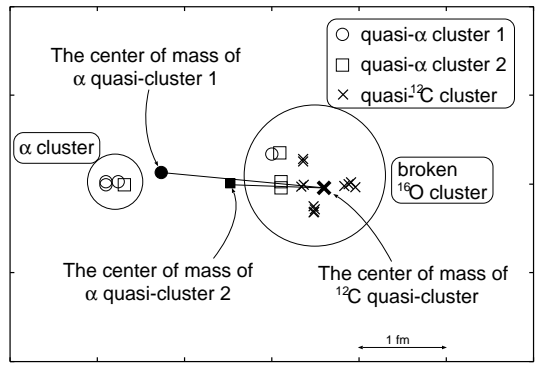}}
\caption{The positions of wave packet centers of nucleons: The wave function forms ``$\alpha$-(broken \ncl{16}{O})'' structure though we constrained two of ``$\alpha$-\ncl{12}{C}'' distances.}
\label{fig:wave_packets_of_exchange}
\end{wrapfigure}
In the case of ``$2\alpha$-\ncl{12}{C}'' wave functions, we constrained the two of ``$\alpha$-\ncl{12}{C}'' distances, $d_1$ and $d_2$, not ``$\alpha$-$\alpha$'' distance. The ``$\alpha$-$\alpha$'' distance was optimized by energy variation. The \ncl{12}{C} quasi-cluster does not need to have a compact 3$\alpha$ structure, but can have different structure for different values of $d$-constraint parameters $d_1$ and $d_2$. An important thing to pay attention is that nucleons belonging to different quasi-clusters can belong to the same cluster after energy variation. See Fig.\ref{fig:wave_packets_of_exchange}, which is the result of calculation with the constraint of $d_1=1\ {\rm fm},\  d_2=2\ {\rm fm}$. It can be seen that there are one $\alpha$ cluster and one broken \ncl{16}{O} cluster. The reason is that three nucleons of the first $\alpha$ quasi-cluster and one nucleon of the second $\alpha$ quasi-cluster make an $\alpha$ cluster  because $d$-constraint does not constrain the position of each nucleon.

As we mentioned in the introduction, only by the $\beta$-constraint, it is sometimes difficult to create the cluster-like wave functions especially for medium-weight nuclei. By the $d$-constraint, we can now create the wave functions rather easy that have cluster aspects. 

\section{The Wave Functions Used in the Present GCM Calculation for \ncl{20}{Ne}}
\label{sec:wave_functions}
We created \ncl{20}{Ne} wave functions of various types of structures which include mean-field-type and $\alpha$-\ncl{16}{O} structure by $\beta$-constraint, and $2\alpha$-\ncl{12}{C} and $\alpha$-\ncl{16}{O} structures by $d$-constraint. 
Only by $\beta$-constraint, we could not create the wave functions of ``$2\alpha$-\ncl{12}{C}'' structures. 
The reason is that $2\alpha$-\ncl{12}{C} states have higher energy than mean-field-type states in small deformation and than $\alpha$-\ncl{16}{O} states in large deformation. 
In the GCM calculation, we combined linearly these \ncl{20}{Ne} wave functions.

\begin{wrapfigure}{r}{8cm}
\centerline{\includegraphics[width=7.5cm]{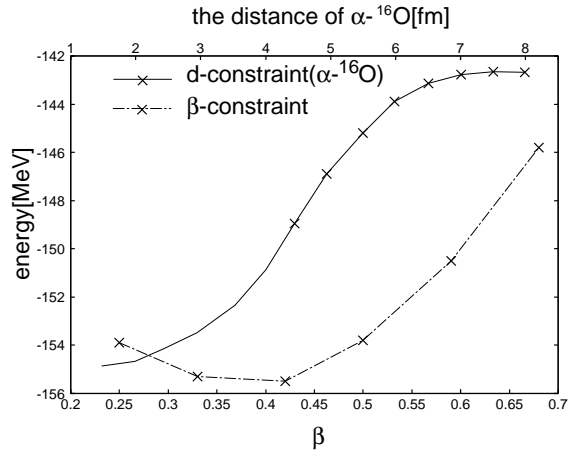}}
\caption{The energy surface of wave functions created by $\beta$-constraint and that of $\alpha$-\ncl{16}{O} wave functions created by $d$-constraint: Upper scale is for $d$-constraint, and lower scale is for $\beta$-constraint. Upper and lower scales have nothing with each other.}
\label{fig:energy_surface}
\end{wrapfigure}
The energy surfaces of the wave functions created by $\beta$-constraint and by the ``$\alpha$-\ncl{16}{O}'' distance constraint are shown in Fig.\ref{fig:energy_surface}. In GCM calculation, we used $\times$-marked-wave functions in the figure. We did not use ``$\alpha$-\ncl{16}{O}'' wave functions with constrained $d\le 4.0\ \rm fm$, because for $d\leq 4.0$ fm nucleon wave packets were deformed and the wave functions were not different so much from the wave functions created by $\beta$-constraint. The nucleon wave packets of ``$\alpha$-\ncl{16}{O}'' wave functions with constrained $d > 4.0\ \rm fm$ are spherical and the wave functions are similar to the Brink model wave functions. As for the wave functions created by $\beta$-constraint, there is detailed explanation in Ref.\citen{kimura}.

\begin{table}[bt]
\caption{The ``$2\alpha$-\ncl{12}{C}'' wave functions used in GCM : $d_{1,2}$ are the distances between the first and the second quasi-$\alpha$ cluster and \ncl{12}{C} cluster. $d_3$ is the distance between two quasi-$\alpha$ clusters, $E_{0^+}$ is the energy of the projected $J^\pi = 0^+$ state in MeV, and $V_{LS}$ is the spin-orbit part of $E_{0^+}$ in MeV.}
\label{tab:d_2alpha-12C}
\begin{center}
\begin{tabular}{clccccccl}
\hline \hline
		&		& $d_1$	& $d_2$	& $d_3$	& $E_{0^+}$	& $V_{LS}$	& $\beta$	& 	 \\ \hline
$d_1=d_2$	& spherical	& 3.09	& 3.15	& 2.14	& $-152.4$	& $-10.4$	& 0.36		&	 \\
		& wave		& 3.20	& 3.24	& 1.75	& $-151.7$	& $-11.6$	& 0.37		&	 \\
		& packets	& 3.32	& 3.33	& 1.94	& $-150.8$	& $-11.9$	& 0.40		&	 \\ \cline{2-9}
		& deformed	& 3.13	& 3.13	& 1.75	& $-153.7$	& $\  -6.1$	& 0.55		&	 \\
		& wave		& 3.23	& 3.24	& 1.82	& $-152.7$	& $\  -7.2$	& 0.55		&	 \\
		& packets	& 3.44	& 3.45	& 1.96	& $-150.8$	& $\  -8.8$	& 0.57		&	 \\ \hline
$d_1<d_2$	& deformed	& 0.96	& 1.97	& 1.05	& $-158.2$	& $\  -4.9$	& 0.43		& $\alpha$-(broken \ncl{16}{O})\\
		& wave		& 1.93	& 2.45	& 0.70	& $-158.6$	& $\  -4.8$	& 0.44		& (broken \ncl{8}{Be})-\ncl{12}{C}\\
		& packets	& 1.88	& 2.96	& 1.74	& $-158.3$	& $\  -4.5$	& 0.45		&	 \\ \hline
\end{tabular}
\end{center}
\end{table}
The characteristics of ``$2\alpha$-\ncl{12}{C}'' wave functions created by $d$-constraint are given in Table \ref{tab:d_2alpha-12C}. 
We adopted two types of constraint, $d_1=d_2$ and $d_1<d_2$. 
We did not constrain the distance $d_3$ between two quasi-$\alpha$ clusters in nether cases. 
As we mentioned in \S\ref{sec:energy_variation}, in the case of $d$-constraint calculation, the resultant distance value after energy variation sometimes differs slightly from the aimed distance value. 
For example, in the first three lines of Table \ref{tab:d_2alpha-12C}, the aimed constraint of $d_1=d_2$ is slightly violated in the resultant values of $d_1$ and $d_2$, and also the aimed values of $d_1=d_2=3.2,\  3.3,$ and $3.4\ \rm fm$ for the first, second, and third lines, are slightly changed in the resultant values of $d_1$ and $d_2$. 
Among the wave functions obtained with the constraint of $d_1=d_2$, we adopted the wave functions that have two $\alpha$ clusters and a \ncl{12}{C} cluster. 
Only when we constrained $d_1=d_2\simeq 3.2\ \rm fm$, we got such wave functions. 
Those wave functions are divided into two types of nucleon wave packets, spherical and deformed. 
One of the largest difference between these wave functions is spin-orbit energy. 
The wave functions with spherical nucleon wave packets have larger spin-orbit energy than those with deformed nucleon wave packets. 
It is because the \ncl{12}{C} cluster of the spherical wave packets is closer to the ground state whose dominant component is the $0p_{\frac{3}{2}}$ subshell closed structure in the case of the Gogny D1S force. 
But the \ncl{12}{C} cluster of the deformed wave packets has a triangular configuration of three prolate deformed $\alpha$ in \ncl{12}{C} cluster which has small spin-orbit energy. 
As the three $\alpha$ in the \ncl{12}{C} cluster are deformed, spin-orbit energy is not zero but its magnitude is very small compared with the spin-orbit energy of the ground state of an isolated \ncl{12}{C} in the Gogny D1S force which amounts to $-17\rm MeV$. 
As for $d_1<d_2$, we adopted the wave functions that have the lowest intrinsic energy within the same condition of constraint. 
The $\alpha$-(broken \ncl{16}{O}) wave function has an $\alpha$ cluster and a broken \ncl{16}{O} cluster by the reason explained in \S\ref{sec:d-constraint}. 
The (broken \ncl{8}{Be})-\ncl{12}{C} wave function has \ncl{12}{C} cluster and eight nucleons not clustering. 
The last wave function obtained with $d_1<d_2$ has a ($2\alpha$-\ncl{12}{C})-like structure, but here also the nucleon wave packets are largely deformed and the \ncl{12}{C} cluster has a three deformed-$\alpha$ structure. 

We expected at first that the $\beta$-constraint wave functions in GCM calculation would mainly contribute in reproducing $K^\pi=0_1^+,0_4^+$ bands, the $d_1=d_2$ ``$2\alpha$-\ncl{12}{C}'' wave functions in reproducing $K^\pi =0_3^+$ band, and the $d_1<d_2$ ``$2\alpha$-\ncl{12}{C}'' wave functions in reproducing $K^\pi =0_2^+$ band. However, as an actual result, the $d_1=d_2$ ``$2\alpha$-\ncl{12}{C}'' wave functions with deformed wave packets and the $d_1<d_2$ ``$2\alpha$-\ncl{12}{C}'' wave functions became mainly the components of $K^\pi =0_1^+$ band.

\section{Results}
\label{sec:results}
\subsection{The energy spectrum}
\begin{wrapfigure}{r}{10cm}
\centerline{\includegraphics[width=9.5cm]{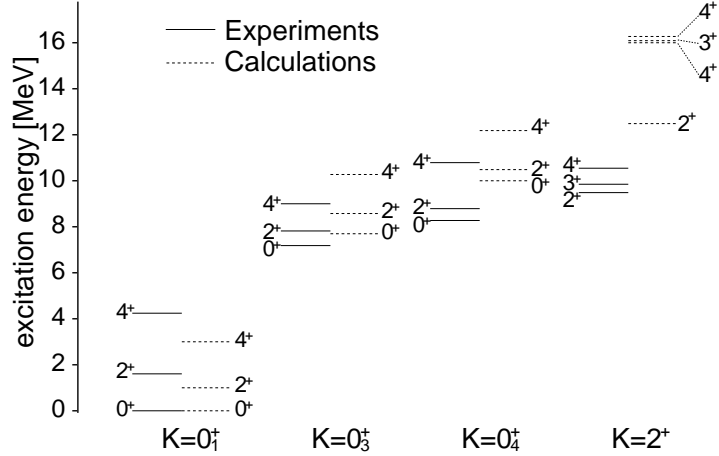}}
\caption{The excitation energies of the low-lying states of \ncl{20}{Ne}}
\label{fig:level}
\end{wrapfigure}
After the angular momentum projection, we have performed the GCM calculation, and obtained the level scheme of \ncl{20}{Ne}(Fig.\ref{fig:level}). 
We have obtained three $K^\pi = 0^+$ bands and a $K^\pi = 2^+$ band.

We consider that the first, second, and third $J^\pi =0^+$ states of our calculations correspond to the observed $J^\pi = 0_1^+\  0_3^+$, and $0_4^+$ states, respectively. The reason is discussed in the next subsection, \S\ref{sec:reason_0_3^+}.

\subsection{Analyzing of the Results of GCM Calculation}
\label{sec:reason_0_3^+}
\subsubsection{Radii, deformations, and harmonic oscillator quantum numbers}
\begin{wraptable}{r}{4.5cm}
\caption{Radii, deformations, and harmonic oscillator quantum numbers: $R$ is r.m.s. radius in fm, $\beta$ is quadrupole deformation parameter, and $N$ is harmonic oscillator quantum number.}
\label{tab:radius_beta_N}
\begin{center}
\begin{tabular}{cccc}
\hline\hline
$J^\pi$	& $R$	& $\beta$	& $N$	\\ \hline
$0_1^+$	& 3.00	& 0.50		& 21.2	\\
$0_3^+$	& 3.22	& 0.69		& 24.8	\\
$0_4^+$	& 3.61	& 0.87		& 31.6	\\ \hline
\end{tabular}
\end{center}
\end{wraptable}
The radii, deformation, and harmonic oscillator number of the $J^\pi =0^+$ states are given in Table \ref{tab:radius_beta_N}. 
The harmonic oscillator number operator, $\hat{N}$, is defined in following way:
\begin{equation}
\hat{N} = \sum_{i=1}^A\left( \frac{\hat{\vect{p}}_i^2}{2m} + \frac{1}{2}m\hat{\vect{r}}_i^2 \right)/ \hbar \omega - \frac{3}{2}A,
\end{equation}
where $A$ is a mass number, $m$ is the nucleon mass, $\hat{\vect{p}}_i$ are  momentum operators, $\hat{\vect{r}}_i$ are position operators, and $\omega$ is determined so that $\left<\hat{N}\right>_{J^\pi =0_1^+}$ takes the lowest value. 

If wave function is of $(sd)^4$ structure, $N$ is equal to 20, and if wave function is of $(p)^{-4}(sd)^8$ structure, $N$ is equal to 24. 
The observed $K^\pi =0_2^+$ band is regarded as having a $(sd)^4$ shell structure, so we consider that the second $J^\pi =0^+$ of our calculation does not correspond to the $J^\pi =0_2^+$ of experiments. 

\newpage
\subsubsection{The components of $\alpha$-\ncl{16}{O}, $2\alpha$-\ncl{12}{C}, and ($\alpha$-\ncl{16}{O})$\cup$($2\alpha$-\ncl{12}{C}) structure}
\begin{wraptable}{r}{7.5cm}
\caption{The components of $\alpha$-\ncl{16}{O}, $2\alpha$-\ncl{12}{C}, and ($\alpha$-\ncl{16}{O})$\cup$($2\alpha$-\ncl{12}{C}) structure: $E$ is total energy in MeV, $c_{\{\alpha\} }$, $c_{\{ 2\alpha\} }$ and $c_{\{\alpha\}\cup\{ 2\alpha\} }$ are the percentages of the component of $\alpha$-\ncl{16}{O}, $2\alpha$-\ncl{12}{C} and ($\alpha$-\ncl{16}{O})$\cup$($2\alpha$-\ncl{12}{C}) structure, respectively.}
\label{tab:components}
\begin{center}
\begin{tabular}{cccccc}
\hline\hline
$J^\pi$	& $E$		& $c_{\{\alpha\} }$	& $c_{\{ 2\alpha\} }$	& $c_{\{\alpha\}\cup\{ 2\alpha\} }$	& $K^\pi$	\\ \hline
$0^+$	& $-160.6$	& 59				& 41				& 78								& $0_1^+$	\\
	& $-152.9$	& 51				& 29				& 90								& $0_3^+$	\\
	& $-150.6$	& 69				& 14				& 92								& $0_4^+$	\\ \hline
$2^+$	& $-159.6$	& 58				& 31				& 72								& $0_1^+$	\\
	& $-152.0$	& 51				& 26				& 83								& $0_3^+$	\\
	& $-150.1$	& 60				& 29				& 93								& $0_4^+$	\\ \hline
\end{tabular}
\end{center}
\end{wraptable}
The amount of the components of $\alpha$-\ncl{16}{O}, $2\alpha$-\ncl{12}{C} and ($\alpha$-\ncl{16}{O})$\cup$($2\alpha$-\ncl{12}{C}) structures are given in Table \ref{tab:components}. 
The amount of the component of $\{ X\}$ structure, $c_{\{ X\} }$, is defined as
\begin{equation}
c_{\{ X\} } = \frac{\bra{\rm\Phi}P_X\ket{\rm\Phi}}{\norm{\rm\Phi}},
\end{equation}
where $P_X$ is the projection operator which projects out the $\{ X\}$ component from the $\rm\Phi$, and $P_X$ is defined in Appendix.
As the basis states which span the $\alpha$-\ncl{16}{O} functional subspace, we adopted the eight wave functions created by the $d$-constraint on the ``$\alpha$-\ncl{16}{O}'' distance $d=4.5,\  5.0,\  5.5,\  6.0,\  6.5,\  7.0,\  7.5$ and $8.0$ fm, as the basis states which span the $2\alpha$-\ncl{12}{C}, we adopted three wave functions created by the $d$-constraint on the two of ``$\alpha$-\ncl{12}{C}'' distance, $d_{1,2}$, $d_1=d_2=3.2,\  3.3$ and $3.4$ fm that have two $\alpha$ clusters and a \ncl{12}{C} cluster with spherical nucleon wave packets, and as for the combined cluster subspace of ($\alpha$-\ncl{16}{O})$\cup$($2\alpha$-\ncl{12}{C}), we adopted both basis wave functions.

When we compare the second and third $J^\pi =0^+$ of our calculation, we find the following points. 
The component of $2\alpha$-\ncl{12}{C} in the second $J^\pi =0^+$ is larger than that of third $J^\pi =0^+$, and the component of $\alpha$-\ncl{16}{O} in the second $J^\pi =0^+$ is smaller than that of third $J^\pi=0^+$. 
This leads that the second $J^\pi =0^+$ of our calculation corresponds to the observed $J^\pi =0_3^+$ and the third $J^\pi =0^+$ of our calculation corresponds to the observed $J^\pi =0_4^+$ because as explained in \S\ref{sec:intro} the observed $K^\pi =0_3^+$ states have been considered to have $2\alpha$-\ncl{12}{C} structure and the observed $K^\pi = 0_4^+$ states have long been known to have $\alpha$-\ncl{16}{O} structure. 

The percentages of the $\alpha$-\ncl{16}{O} and $2\alpha$-\ncl{12}{C} components contained in the calculated second $J^\pi = 2^+$ state ($2_2^+$) are similar to those of the calculated second $J^\pi = 0^+$ state ($0_2^+$), but those of the calculated third $2^+$ state ($2_3^+$) are slightly different from those of the calculated third $0^+$ state ($0_3^+$), namely the $\alpha$-\ncl{16}{O} component slightly decreases from $0_3^+$ to $2_3^+$ while the $2\alpha$-\ncl{12}{C} component slightly increases from $0_3^+$ to $2_3^+$. 
Although the $2\alpha$-\ncl{12}{C} component of $2_3^+$ is now comparable to or even slightly larger than that of $2_2^+$, we classify the calculated $2_2^+$ and $2_3^+$ states to the $K^\pi = 0_3^+$ and $0_4^+$ bands, respectively, because the $\alpha$-\ncl{16}{O} component of $2_3^+$ is larger than that of $2_2^+$ which, we consider, means that $2_3^+$ corresponds to the $K^\pi = 0_4^+$ band member. 

We can say that the most importance character of the $K^\pi = 0_4^+$ band is that it has a prominent $\alpha$-\ncl{16}{O} cluster structure and almost all the calculations have never failed in reproducing the $K^\pi = 0_4^+$ band as far as the model space contains the $\alpha$-\ncl{16}{O} cluster subspace. 
Therefore we think that our assignment of the calculated $0_3^+$ and $2_3^+$ states to the $K^\pi = 0_4^+$ band members is rather reliable. 

On the other hand, we think the our present treatment of the $2\alpha$-\ncl{12}{C} subspace is not sufficient, because, as we explained in \S\ref{sec:wave_functions}, the GCM basis states which can be said to have $2\alpha$-\ncl{12}{C} structure are only three states obtained by the $d$-constraint with $d_1=d_2=3.2,\  3.3,$ and $3.4\ \rm fm$. 
If we include more basis states with $2\alpha$-\ncl{12}{C} structure, the percentages of the $2\alpha$-\ncl{12}{C} component in the calculated $0_2^+$ and $2_2^+$ will increase while those in the calculated $0_3^+$ and $2_3^+$ will decrease. 

\subsubsection{The components of the binding energy}
\begin{wraptable}{r}{8.5cm}
\begin{center}
\caption{The components of the binding energy: $E$ is total energy, $T$ is kinetic energy, $V_2$ is 2-body energy, $V_{LS}$ is spin-orbit energy, $V_{\rm C}$ is Coulomb energy, and $V_{\rm g}$ is density dependent energy in MeV}
\begin{tabular}{ccccccccc}
\hline\hline
$J^\pi$	& $E$		& $T$	& $V_2$		& $V_{LS}$	& $V_{\rm C}$	& $V_{\rm g}$ \\
\hline
$0_1^+$ & $-160.6$ 	& 283.5 & $-892.0$	& $-5.7$	& 19.1		& 434.5 \\
$0_3^+$ & $-152.9$ 	& 280.2 & $-866.2$	& $-5.0$	& 18.6		& 419.5 \\
$0_4^+$ & $-150.6$ 	& 273.3 & $-847.9$	& $-3.2$	& 17.8		& 409.3 \\
\hline
\end{tabular}
\label{tab:energy_components}
\end{center}
\end{wraptable}
The decomposition of the calculated binding energy is given in Table \ref{tab:energy_components}. 
$J^\pi =0_4^+$ state has less spin-orbit energy than $J^\pi =0_1^+$ and $0_3^+$ states. 
It is because $J^\pi =0_4^+$ state has more $\alpha$-\ncl{16}{O} component, of which spin-orbit energy is nearly equal to zero, than $J^\pi =0_1^+$ and $0_3^+$ states.

\section{Summary}
\label{sec:summary}
We have proposed a new constraint of clustering for the AMD calculation, which we call {\it $d$-constraint}. 
By the $d$-constraint, we can create rather easily the wave functions that have cluster aspect. 
It is very helpful and important for the AMD study because in AMD approach we sometimes find difficulty in getting the wave functions with some specific cluster structure. 

We have applied the $d$-constraint to \ncl{20}{Ne} nucleus. 
We could create rather easily $2\alpha$-\ncl{12}{C} wave functions which were not easy to get in the AMD approach. 
Using these wave functions, we could reproduce $K^\pi = 0_3^+$ band which contains the $2\alpha$-\ncl{12}{C} structure as an important component.
It should be noted that even the existence of the $K^\pi = 0_3^+$ band has not been shown before not only by AMD but also by the Hartree-Fock approach. 

Now we can create rather easily the cluster-like wave functions for medium-weight nuclei. 
We expect that our new constraint is powerful for the study of the coexistence of the cluster structure and mean-field structure in $sd$ and $pf$-shell nuclei and the study of the relation between superdeformation and cluster structure.

\section*{Acknowledgements}
Many of the computations has been done on a supercomputer SX-5 at the Research Center for Nuclear Physics, Osaka University (RCNP).

\appendix
\section{}
Suppose the states $\ket{\phi_i}$ span the functional space $\{ X\}$. 
Their overlap matrix $B_{ij}$ is defined as
\begin{equation}
B_{ij} = \inpro{\phi_i}{\phi_j}.
\end{equation}
The orthonormalized basis wave functions of the $\{ X\}$, $\ket{\tilde{\phi}_\alpha}$, are given by the linear combination of the $\ket{\phi_i}$,
\begin{equation}
\ket{\tilde{\phi}_\alpha} = \frac{1}{\sqrt{\rho_\alpha}}\sum_ic_{i\alpha}\ket{\phi_i},
\end{equation}
where the $\rho_\alpha$ and $c_{i\alpha}$ are the eigenvalues and eigenvectors of the overlap matrix $B_{ij}$, respectively, 
\begin{equation}
\sum_jB_{ij}c_{j\alpha} = \rho_\alpha c_{i\alpha}.
\end{equation}
The eigenvectors $c_{i\alpha}$ form a complete orthonormalized set,
\begin{equation}
\sum_\alpha c_{i\alpha}^*c_{j\alpha} = \delta_{ij};\  \sum_i c_{i\alpha}^*c_{i\beta} = \delta_{\alpha\beta}.
\end{equation}
It is easy to show that \ket{\tilde{\phi}_\alpha} are orthonormal.

Using \ket{\tilde{\phi}_\alpha}, $P_X$ is defined as,
\begin{equation}
P_X = \sum_\alpha \ket{\tilde{\phi}_\alpha}\bra{\tilde{\phi}_\alpha}.
\end{equation}

\end{document}